\documentclass[axioms,article,accept,pdftex,moreauthors]{Definitions/mdpi} 
\firstpage{1} 
\pubvolume{1}
\issuenum{1}
\articlenumber{0}
\pubyear{2023}
\copyrightyear{2023}
\externaleditor{Academic Editor: Oscar Humberto Montiel Ross}
\datereceived{10 February 2023} 
\daterevised{28 March 2023} 
\dateaccepted{28 March 2023} 
\datepublished{31 March 2023} 
\hreflink{https://doi.org/} 

\usepackage{natbib}
\usepackage[export]{adjustbox}
\usepackage{xcolor}

\usepackage{graphicx}	
\usepackage{amsmath}	
\usepackage{subcaption}
\usepackage{amssymb}	

\Title{One-dimensional 
Convolutional Neural Networks for Detecting Transiting~Exoplanets}

\TitleCitation{One-dimensional Convolutional Neural Networks for Detecting Transiting Exoplanets}

\Author{Santiago Iglesias Álvarez $^{1,2}$\orcidA{}, Enrique Díez Alonso $^{1,3}$\orcidC{}, María Luisa Sánchez Rodríguez $^{1,2}$\orcidB{}, Javier~Rodríguez~Rodríguez~$^{1,4}$\orcidD{}, Fernando Sánchez Lasheras $^{1,3}$*\orcidE{} and Francisco Javier de Cos Juez $^{1,4}$\orcidF{}}

\AuthorNames{Santiago Iglesias Álvarez, Enrique Díez Alonso, María Luisa Sánchez, Javier Rodríguez Rodríguez, Fernando Sánchez Lasheras and Francisco Javier de Cos Juez}

\AuthorCitation{Iglesias Álvarez,~S.; Díez Alonso,~E.; Sánchez,~M.L.; Rodríguez Rodríguez,~J.; Sánchez Lasheras,~F.; de Cos Juez,~F.J.}

\address{%
$^{1}$ \quad Instituto Universitario de Ciencias y Tecnologías Espaciales de Asturias (ICTEA), C. Independencia 13, 33004~Oviedo, Spain; iglesiasalvarezsantiago@gmail.com (S.I.Á); diezenrique@uniovi.es (E.D.A.); mlsr@uniovi.es (M.L.S.R.); rodriguezrjavier@uniovi.es (J.R.R.); fjcos@uniovi.es (F.J.d.C.J.)\\
$^{2}$ \quad Departamento de Física, Universidad de Oviedo, 33007 Oviedo, Spain\\
$^{3}$ \quad Departamento de Matemáticas, Facultad de Ciencias, Universidad de Oviedo, 33007 Oviedo, Spain\\
$^{4}$ \quad Departamento de Explotación y Prospección de Minas, Universidad de Oviedo, 33004 Oviedo, Spain\\ }

\corres{Correspondence: sanchezfernando@uniovi.es}

\abstract{The transit method is one of the most relevant exoplanet detection techniques, which consists of detecting periodic eclipses in the light curves of stars. This~is not always easy due to the presence of noise in the light curves, which is induced, for example, by the response of a telescope to stellar flux. For~this reason, we aimed to develop an artificial neural network model that is able to detect these transits in light curves obtained from different telescopes and surveys. We~created artificial light curves with and without transits to try to mimic those expected for the extended mission of the Kepler telescope (K2) in order to train and validate a 1D convolutional neural network model, which was later tested, obtaining an accuracy of 99.02\% and an estimated error (loss function) of 0.03. These results, among others, helped to confirm that the 1D CNN is a good choice for working with non-phased-folded Mandel and Agol light curves with transits. It~also reduces the number of light curves that have to be visually inspected to decide if they present transit-like signals and decreases the time needed for analyzing each (with respect to traditional analysis).}


\begin{document}



\section{Introduction}

After the discovery of the first exoplanetary system in 1992~\citep{1992Natur.355..145W} by using the radial velocity (RV) technique, which consists of the measurement of Doppler shifts in the spectral lines of a star due to the gravitational interaction with an orbiting exoplanet, the detection techniques and facilities have considerably improved, allowing the detection of thousands of planetary systems.

One of the most relevant exoplanet detection techniques is the transit method, which consists of detecting periodic eclipses in the light curves of stars due to the crossing of a planet with the line of sight between a telescope and its host star. When this effect arises, the stellar brightness (transit) decreases. A model that most accurately describes the theoretical transit shape was proposed~\citep{Mandel_2002} (which we refer to as the Mandel and Agol theoretical shape), which models them as an overlap of a spherical and opaque object (planet) above another bright sphere (star) considering the limb-darkening effect, which is an optical effect that makes the star appear less bright on the edges than on the center, as previously described~\citep{2000A&A...363.1081C}. The~first discovery of an exoplanet with this technique was carried out by~\citep{Charbonneau_2000, Henry_2000}, in which an Earth-like exoplanet was discovered orbiting the star HD 209458.

The aim of the Kepler mission~\citep{doi:10.1126/science.1185402}, which was launched in 2009, was to detect Earth-size planets orbiting Sun-like stars and located in their habitability zone (HZ) (The habitability zone is the region around a star where liquid water may exist), through the transits technique. It~was designed considering the wide variety of hosts stars in which exoplanets can be found. In its main mission, it monitored stars in a field of the sky located between the constellations of Cygnus, Lyra, and Draco. The~Kepler mission captured data every 30~min (long cadence) of most of the stars, with the exception of a small group of stars, which were monitored each 2~min (short cadence). The~main mission finished with the discovery 2710 planets~\citep{Akeson_2013}.

The extended mission of Kepler, K2~\citep{Howell_2014} started in 2014 after the break of two of the reaction wheels of the spacecraft. It~captured data from different target fields located throughout the ecliptic. Each field was observed in a campaign of an average duration of 75~days, except the last one (campaign 19), due to the end of the mission when the spacecraft ran out of fuel. An extra campaign (campaign 0) was used as a test. The~stability in the axis without a reaction wheel was achieved by minimizing the solar pressure along this axis. For~that aim, the spacecraft was orientated in the antisolar direction. The~problem was that the radiation pressure over long time scales induced a movement of the telescope, which was corrected by firing its thrusters. This~movement of the telescope results in the stars appearing in different pixels of the detector, which produced a non-stellar-origin trend in the light curves that must be subtracted before analyzing them, and that was one of the main difficulties in obtaining the same precision as in the primary mission. It~also captured data with long and short cadences. The~main goal of the K2 mission was to continue stably operating with the Kepler telescope considering all the limitations induced by the two lost reaction wheels and thus allowing the discovery of more exoplanets. Moreover, the K2~mission allowed the discovery of 543 more exoplanets~\citep{Akeson_2013}. Some examples of K2-confirmed exoplanets have been reported~\citep{2018MNRAS.476L..50D, 2018MNRAS.480L...1D, 2019MNRAS.489.5928D}.

Apart from Kepler, other relevant surveys and telescopes allow the use of the transits technique. Some of them, such as SuperWASP~\citep{Pollacco_2006}, are terrestrial, whereas others, such  as CoRoT~\citep{2009A&A...506..411A}, TESS~\citep{10.1117/1.JATIS.1.1.014003}, and CHEOPS~\citep{2021ExA....51..109B}, are located in space.

Traditional transit analyses are conducted by applying periodogram-like algorithms to the light curves of stars. The~most relevant are box least squares (BLS)~\citep{refId0}, which models the transits as squared boxes, and transit least squares (TLS)~\citep{TLS}, which models them by implementing the theoretical Mandel and Agol shape. All of them highly accurately detect transit-like signals in the light curves and allow the estimation of some parameters related to the planetary system, such as the planet's radius in terms of the radius of its host star, orbital period, etc. However, their computational cost was high, and all the results obtained had to be checked to interpret the presence of transit-like signals in the light curves.

Additionally, the light curves present long-term trends that are caused by stellar variability phenomena, such as stellar rotation~\citep{2017hsa9.conf..502D} or pulsations, or by the extraction of the photometry in the telescope sensor. The~light curves also present Poisson noise, which depends on the sensor accuracy, related to the apparent magnitude of the star (Apparent magnitude, usually referred to as magnitude, is a measure of the brightness of a star observed from the Earth. The~magnitude scale is inverse logarithmic, where the brightest stars have the lowest values). Specifically, the noise is lower in brighter stars. All these effects hinder the prediction of the presence of transit-like signals in the light curves by visual inspection in order to only analyze the ones with these signals with periodogram-like algorithms, thus reducing the computational cost. This~is why artificial intelligence (AI) techniques have started to be introduced as the solution.


Machine learning (ML) is a discipline of AI that is based on different mathematical algorithms that allow computers to identify common features in data. ML techniques overcome the need for human judgment in classifying transit-like signals in candidates, false positives, etc., by using an automatic vetting where the criteria are always the same, which is impossible if this process is simultaneously performed by several researchers. One~of the first trials of this process was the Robovetter project~\citep{Coughlin_2016}, which involved a decision tree trained for refusing false-positive threshold-crossing events (TCEs), which are detected periodic signals that could be consistent with planetary transit signals. Other~researchers started to use ML techniques to provide another approach to this new vetting system. The~Autovetter project~\cite{McCauliff_2015} and Signal Detection using Random-Forest Algorithm (SIDRA)~\cite{2016MNRAS.455..626M} used random forests to classify TCEs based on different transit-like signal features. Other~researchers~\cite{Thompson_2015, 2017MNRAS.465.2634A} took a different approach by using unsupervised ML techniques, specifically Self-Organizing Maps, to group light curves from the Kepler telescope based on their shape, such as the presence of transits in light curves. Moreover, other researchers started to further explore artificial neural networks (ANN) and convolutional neural networks (CNN)~\citep{726791}. The~results provided by~\cite{2018MNRAS.474..478P, Zucker_2018, Ansdell_2018, 2019MNRAS.488.5232C, Shallue_2018, 2021arXiv210506292V,1991Ap&SS.181..313D}, among others, show that CNNs were a better choice than the previous ML techniques for vetting planetary candidates extracted from transit-like signals in different light curves. In the first two studies~\cite{2018MNRAS.474..478P, Zucker_2018}, the authors used simulated light curves as the inputs of their CNN, whereas others~\cite{Shallue_2018} used human-vetted Kepler TCEs.

In this paper, we present a 1D convolutional neural network (1D CNN) model that we trained, validated, and tested with simulated detrended light curves, i.e., light curves after subtracting long-term-trends, the aim of which was to mimic the ones expected for the K2 mission, according to the duration, noise, and data collection cadence. Previous researchers~\cite{2018MNRAS.474..478P, Zucker_2018} explored the ability of different 1D CNN models to learn from simulated light curves. The~main difference of this method from that previously reported~\cite{2018MNRAS.474..478P} is that the simulated data in the previous study had a maximum time domain of 6 h, so their training dataset appears as phase-folded light curves (when they have transits), whereas our simulated light curve time domain spreads to 75~days (the mean theoretical duration of K2 light curves). In contrast, Zucker et al.~\cite{Zucker_2018} injected trapezoidal transits that were different from the theoretical shape of transits proposed by Mandel and Agol models. Additionally, they used a five-minute cadence of data, which, in reality, is not common among current telescopes. As~CNNs have a high dependence on the shape of the input data, because the transformation that is produced by convolution kernels to the data in each layer depends on the initial shape, we thought that it was better to focus our efforts on simulating light curves from only one survey (in our case, K2).

The remainder of this article is structured as follows: In Section \ref{sec: CNN}, we explain the theory related to convolutional neural networks (CNNs). In Section \ref{sec: model}, we describe our 1D convolutional neural network model. In Section \ref{sec: train_test}, we describe the simulation and verification of the datasets that were used to train and test our CNN model. In Section \ref{sec: train_results}, we describe the training process, the validation, the test,  the results obtained, and their discussion. Finally, in Section \ref{sec: conclusions}, we summarize the study.

\section{Materials and Methods}
\subsection{Convolutional Neural Networks (CNNs)\label{sec: CNN}}
Artificial neural networks (ANNs) try to mimic the human brain and how its neurons learn. They are formed by interconnected layers composed of artificial nodes or neurons. Deep learning (DL) is a branch of machine learning (ML) that can resolve different tasks that cannot be achieved with traditional ML techniques~\citep{SIG-039}. As~referenced in its name, DL consists of an architecture composed of multiple layers with nonlinear activation functions, which are interconnected by neurons. All the layers except for the first and last layers are known as deep layers. An ANN can be trained with different algorithms; in our case, we used backpropagation~\citep{6795724}.

Convolutional neural networks (CNNs) are a type of ANN that have performed well in pattern recognition~\citep{8308186}. Two of the most relevant characteristics of CNNs are that they do not consider spatially dependent features and that the features picked up by a CNN from the input data are more abstract in the deeper layers~\cite{726791}. The~name CNN makes reference to the most relevant computational operation applied by this kind of ANN, the convolution~\citep{Goodfellow-et-al-2016}, which is a type of mathematical operation that involves two functions (f and g), resulting in another function (h), which expresses how one of the input functions modifies the other~\citep{doi:10.1126/science.123.3195.512.c}:
\vspace{-6pt}
\begin{equation}
h\equiv (f*g)(t):= \int_{-\infty}^{\infty} f(\tau)g(t-\tau)d\tau
\label{eqn: convolution}
\end{equation}

If we focus on a discrete situation~\citep{damelin_miller}, where complex-valued functions are considered, Equation (\ref{eqn: convolution}) could be expressed as:
\vspace{-6pt}
\begin{equation}
(f*g)[n]= \sum^{+\infty}_{m=-\infty} f[m]g[n-m]
\end{equation}

\begin{figure}[H]
\hspace{-30pt}
\includegraphics[width=9.5 cm]{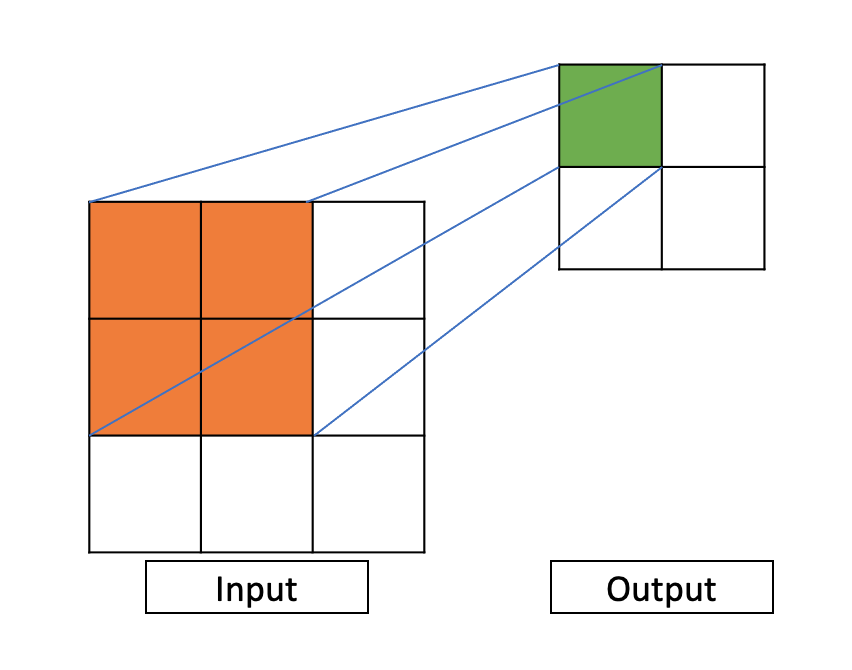}
\caption{Example of how a convolutional filter operates in a fictitious 2D situation. In the input image, in orange, the filter is shown (in this case, a 2~$\times$~2 filter); in the output image, in green, the resulting pixel obtained after applying the filter is shown. As~shown, the data collected from a 2~$\times$~2~pixels cell resulted in only one pixel in the output. This~procedure was repeated through all the rows and columns by moving the filter, transforming a 3~$\times$~3 image into a 2~$\times$~2.}
\label{fig: filters}
\end{figure}

This operation allows the CNN to filter the data and extract features based on the similarities with the applied filters. Figure \ref{fig: filters} shows an example of how a convolutional filter operates in a fictitious 2D situation. We emphasize that we used a 1D CNN model in our study, but a 2D example would be more intuitive and easier to understand.
\vspace{-6pt}

\subsection{Our 1D Convolutional Neural Network Model}\label{sec: model}

For creating and training the ANN model, we used Keras~\citep{chollet2015keras}, which is an open-source Python library for artificial neural networks. It~works as an interface for TensorFlow~\citep{tensorflow2015-whitepaper}.

Our ANN consists of a 1D CNN, which has a five-layer convolutional part and a three-layer multilayer perceptron (MLP) part; in Keras, these layers are known as Dense. The~first convolutional layer (CL) has 16 convolutional kernels (hereafter referred to as kernels) with a filter size of 4 and 2 strides; the second one has 32 kernels with a filter size of 6 and 2 strides; the third one has 64 kernels, with a filter size of 10 and 2 strides; the fourth one has 16 kernels, with a filter size of 18 and 2 strides; and the last one has 16 kernels, with a filter size of 4 and 2 strides. All the convolutional kernels have the padding set to 'same', which forces the output and input feature maps to be the same (assuming a stride of 1), by adding the needed number of 0. Furthermore, we adopted the exponential linear unit (ELU)~\citep{2015arXiv151107289C} as the activation function:

\begin{equation}
ELU(x)= \left\{
\begin{array}{lr}
x & if\: x>0\\
\alpha \left( exp(x)-1 \right) & if\: x\leq 0
\end{array}
\right.
\label{eqn: elu}
\end{equation}

\textls[-15]{This activation function was selected after running the training process (detailed in Section~\ref{sec: train_results}) with the activation functions ELU, ReLU~\citep{2018arXiv180308375A}, hyperbolic tangent, and Sigmoid~\citep{NARAYAN199769}, obtaining the lowest loss function values and the most stable training with ELU.
Then,~we applied a Flatten layer to the output of the convolutional part of our model before connecting it to the MLP part. In this part of the CNN model, the first two Dense layers have 128 neurons and are connected through a Dropout layer with a value of 0.15, which means that 15\% of the neurons of the first Dense layer are disconnected. We~made this decision because of the high number of trainable parameters present in the first Dense layer. These two layers have ELU as the activation function. The~last layer is a two-neuron Dense layer, activated by a SoftMax~\citep{NIPS1989_0336dcba} function. The~standard SoftMax function $\sigma: \mathbb R^K\rightarrow (0,1)^K$ is expressed as (if $K \geq 1$): }

\begin{equation}
\sigma(z)_i = \frac{e^{\beta z_i}}{\sum_{j=1}^K e^{\beta z_j}}
\label{eqn: softmax}
\end{equation}
where $j= 1, ..., K$ and $z= (z_1, ..., z_K) \in \mathbb R^K $.

\begin{figure}[H]
\hspace{-6pt}
\includegraphics[width=0.7\linewidth]{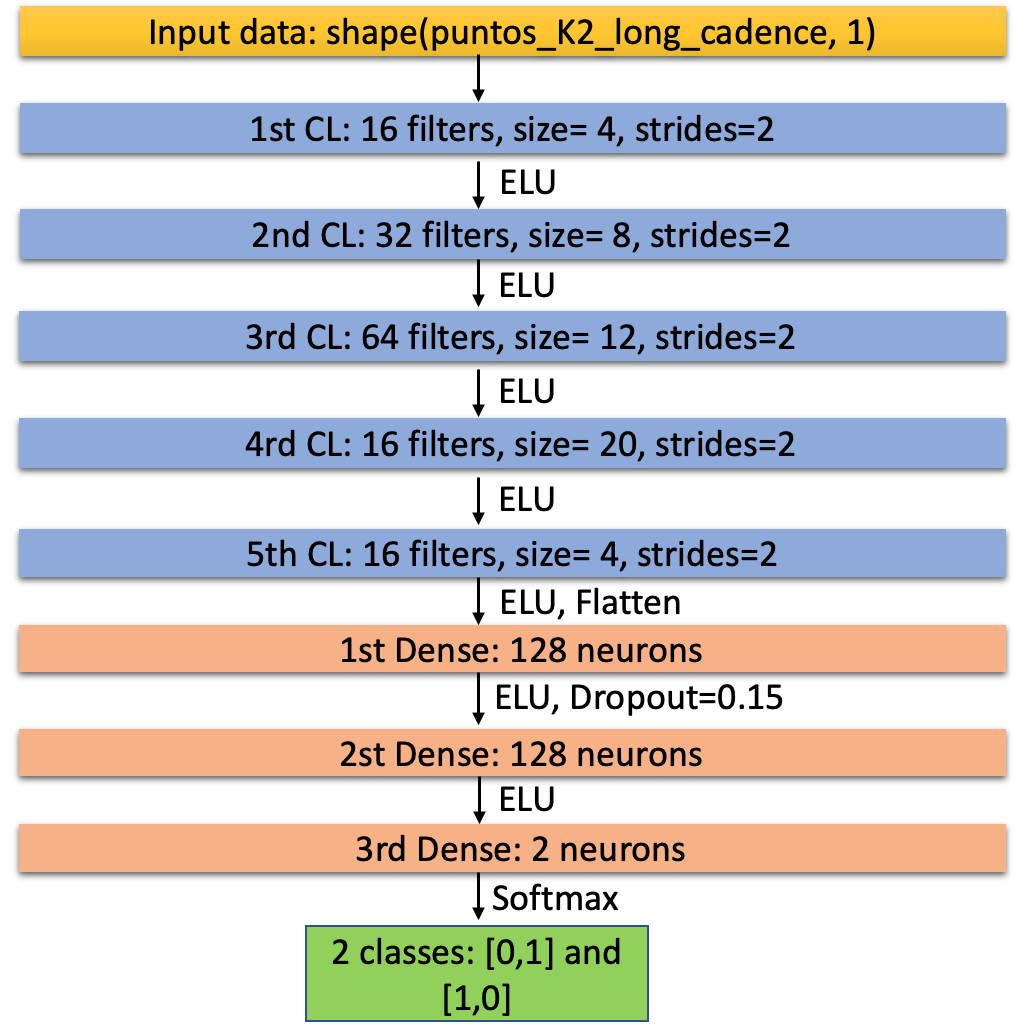}
\caption{Graphic model of our CNN. In yellow, the input data shape; in blue, convolutional layers; in salmon, dense layers; in green, the output of the neural network. The~lines between the layers indicate the activation function, the Dropout, and the Flatten layer, when appropriate.}
\label{fig: CNN_model}
\end{figure}

The choice of two neurons in the last Dense layer was motivated by the fact that in our datasets (Section \ref{sec: train_test}), we have light curves with transits and without transits, i.e., two classes. Thus, we wanted the output of the CNN to be only (1,0) or (0,1), depending on whether transits were present or not. In other words, we aimed to create a CNN classifier that could help us to reduce the amount of time needed to know if a light curve presents transit-like signals or not. A diagram of our CNN model is shown in Figure \ref{fig: CNN_model}.

\subsection{Simulated K2 Light Curves: Training and Testing Datasets} \label{sec: train_test}
\textls[-25]{The theoretical shape of transits is described by Mandel and Agol models~\citep{Mandel_2002}. Without~considering the limb-darkening effect, the flux that reaches the telescope can be described as:}
\begin{equation}
F(a[r_\star], r_p[r_\star])=1-\lambda(a[r_\star], r_p[r_\star])
\label{eqn: mandel_1}
\end{equation}
where $r_p[r_\star]$ is the planet-to-star radius ratio, $a[r_\star]$ is the ratio between the semimajor axis of the orbit and the host star radius, and $\lambda(a[r_\star], r_p[r_\star])$ is the amount of flux blocked by the exoplanet. On the contrary, when this effect is considered, the amount of flux that reaches the telescope is~\citep{Mandel_2002}:
\vspace{-6pt}
\begin{equation}
F(a[r_\star], r_p[r_\star])=\left[ \int^1_0 dr 2 r I(r)\right]^{-1} \int^1_0drI(r)\frac{d[\tilde{F}(a[r_\star]/r, r_p[r_\star]/r])}{dr}
\label{eqn: mandel_2}
\end{equation}
\textls[-15]{where $\tilde{F}(a[r_\star]/r, r_p[r_\star]/r])$ is the non-limb-darkening transit shape described in Equation~(\ref{eqn: mandel_1}). The~limb-darkening effect is parameterized by the term $I(r)$, which computes the intensity emitted by each point of the star. The~intensity of the star considering the nonlinear limb-darkening effect proposed by~\cite{2000A&A...363.1081C} is:}
\vspace{-6pt}
\begin{equation}
I(r)=1-\sum^4_{m=1} c_m (1-\sqrt{1-r^2}^{m/2})
\label{eqn: claret}
\end{equation}

\begin{figure}[H]
\includegraphics[width=0.7\linewidth]{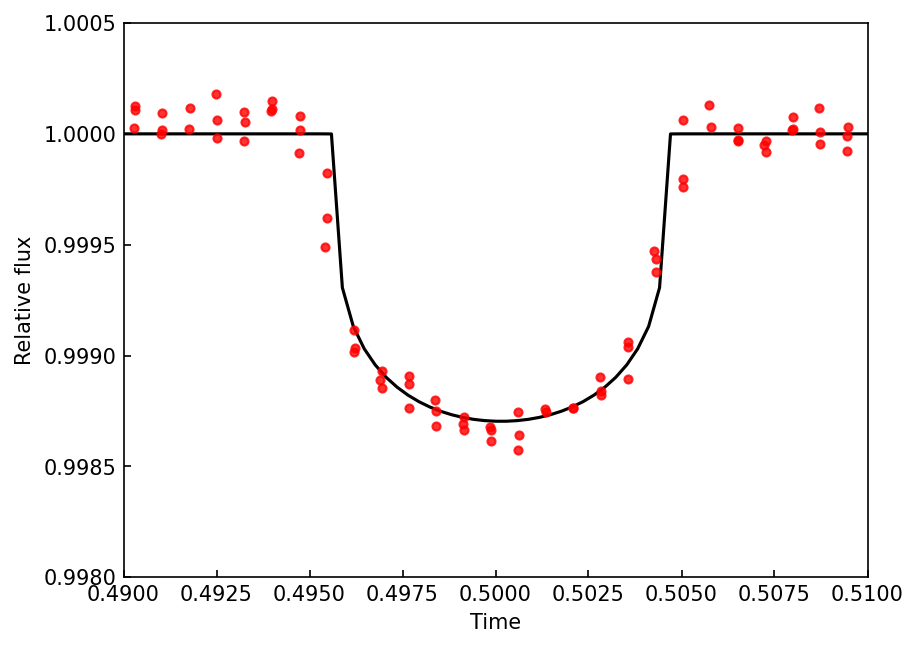}
\caption{Mandel and Agol shape of the transits of K2-110 b. Red dots are the phased-folded light curve, and the black line is the Mandel and Agol model. Computed with TLS.}
\label{fig: model}
\end{figure}

This effect produces a transit shape, which was initially considered a trapezoid, that is rounded, as shown in Figure \ref{fig: model}, where the Mandel and Agol computed model of the light curve from K2-110~\citep{k2-110b} is plotted upon the phase-folded light curve, which consists of folding the light curve upon the epoch, which is the time at which a transit used as a reference (usually the first one) takes place. 

\textls[-15]{For training, validating, and testing our 1D CNN model (see Section \ref{sec: model}), we simulated the light curves most similar to those expected for the K2 mission, so we chose a data-taken cadence of 30~min (long cadence) and a duration of 75~days per light curve,  which are the mean cadence and average temporal span of each campaign of the K2 mission, respectively. This~means that each light curve had a length of 3600 points. For~computing them, we used batman~\citep{Kreidberg_2015}, which is a Python package that models light curves with transits considering the theoretical Mandel and Agol~\citep{Mandel_2002} transit shape. Specifically, it created an ideal detrended light curve, which is the result of computing the point-to-point division of the flux and long-term trend. We~only added  transit-like signals from one simulated exoplanet because the aim of our 1D CNN is to detect if transits-like signals exist or not, so we did not need to train it with simulated systems with two or more transiting exoplanets. In addition, we only needed the flux of each simulated light curve because, as our neural network model is a 1D CNN, it does not consider the time in which transits occur. Instead, it learns their shape.}

The parameters needed for computing a detrended light curve with transits are: the epoch ($t_0$); the orbital period (P) of the simulated planetary system (in~days); the planet-to-star radius ratio ($r_p[r_\star]$), which is the ratio between the planet radius and the host star radius (dimensionless quantity); the ratio between the semimajor axis of the orbit and the host star radius ($a[r_\star]$) (dimensionless quantity); the angle of inclination of the orbital plane ($i$) (in this study, we set this value to 90$^\circ$, which means that the planet crosses the center of the star, which is its brightest part). This~means that the transits are as deep as possible; the limb-darkening coefficients are $u_1$ and $u_2$~\citep{1979ApJS...40....1K, 1985A&AS...60..471W, 2000A&A...363.1081C}. However, not all these parameters are independent. For~example, the orbital period and semimajor axis of the orbit are related through Kepler's third law (Equation (\ref{eqn: kepler})), assuming it is a circular orbit. Thus, we cannot choose both values as random, so we chose the value of the orbital period as a random number between 1~day and a half of the maximum duration of the light curve (75~days) because we wanted at least two transits in the light curves in order to increase the accuracy of transit detection. Additionally, this equation shows a relationship between the host star mass and the semimajor axis of the orbit, and Equation (\ref{eqn: mass-radius})~\citep{1991Ap&SS.181..313D}, for stars in the main sequence (The main sequence (MS) is a continuous region in the Hertzsprung--Russell (HR) diagram, which is a graphic that shows the relationship between the color index (a formula that expresses the color of a star and is related to the temperature) and brightness of stars. These stars fuse hydrogen into helium), shows that a relationship exists between stellar mass and stellar radius. Planetary statistics~\citep{Akeson_2013} show that it is not common to find Jupiter-like exoplanets (usually known as Jupiter or hot Jupiter when $a<0.1 AU$) orbiting low-mass stars (as red dwarfs); thus, picking $r_p[r_\star]$ as a no-limit random value is not a suitable choice. Instead, we chose a value of $r_p[r_\star]\in[0.01, 0.05]$ if $M_\star< 0.75 M_\odot$, which means that a planet orbiting a low-mass star (such as red dwarf) could not be larger than a Neptune-like exoplanet, and $r_p[r_\star]\in[0.01, 0.1]$ if $M_\star\geq 0.75 M_\odot$, which means that the candidate could be as large as a Jupiter-like exoplanet. The~host star mass is chosen as a random value considering the stellar mass distribution in the Solar neighborhood~\citep{2006SerAJ.172...17N}; the epoch is chosen as a random number between 0 and the value of orbital period and the limb-darkening coefficients are computed as random numbers between 0 and 1.
\vspace{-6pt}
\begin{equation}
P^2\sim \frac{4\cdot\pi}{G\cdot M_\star}\cdot a^3 \: (assuming \: M_\star >> M_p)
\label{eqn: kepler}
\end{equation}
\vspace{-6pt}
\begin{equation}
R_\star[R_\odot]\approx M_\star[M_\odot]^{0.8}
\label{eqn: mass-radius}
\end{equation}

The flux of the simulated light curves is expected to be equal to 1 when the transits do not arise and a value lower to 1 (following the Mandel and Agol theoretical shape) when transits arise. The~next step is adding noise to the light curves because one of the most relevant difficulties related to the transit technique is owing to the noise present in the light curves, which is basically due to the response of the telescope to the flux received (which depends on the stellar magnitude). Other~difficulties, for example, come from the inclination angle of the orbital plane (i). If the value of this angle is similar to 0, then transits arise near the limb of the host star, so their size could be underestimated or, directly, they are not detectable. Additionally, a low planet-to-radius ratio can result in the impossibility of detecting transits that are mixed with stellar noise. In addition, the presence of spots in a star can lead to changes in transit shape.

For adding noise to light curves, we used a Gaussian distribution with a mean value of 0 and a standard deviation obtained from~\cite{Koch_2010}. The~values of the stellar magnitudes were obtained as random numbers between 10 and 18 because we did not want light curves with low noise levels, something that can occur with the brightest stars. Thus, we created Gaussian noise vectors with the same length as the light curves computed with batman. As~the mean of the noise vector is 0, we added both vectors (light curve and noise), thus obtaining noisy K2-simulated light curves. As~a transit is decreasing in a light curve, we clipped the flux to $1 \sigma$ above 1.0 to delete all the possible outliers induced by the Gaussian-added noise; we thus prevent our CNN model from learning from them.

\begin{figure}[H]

\includegraphics[width=12.5cm]{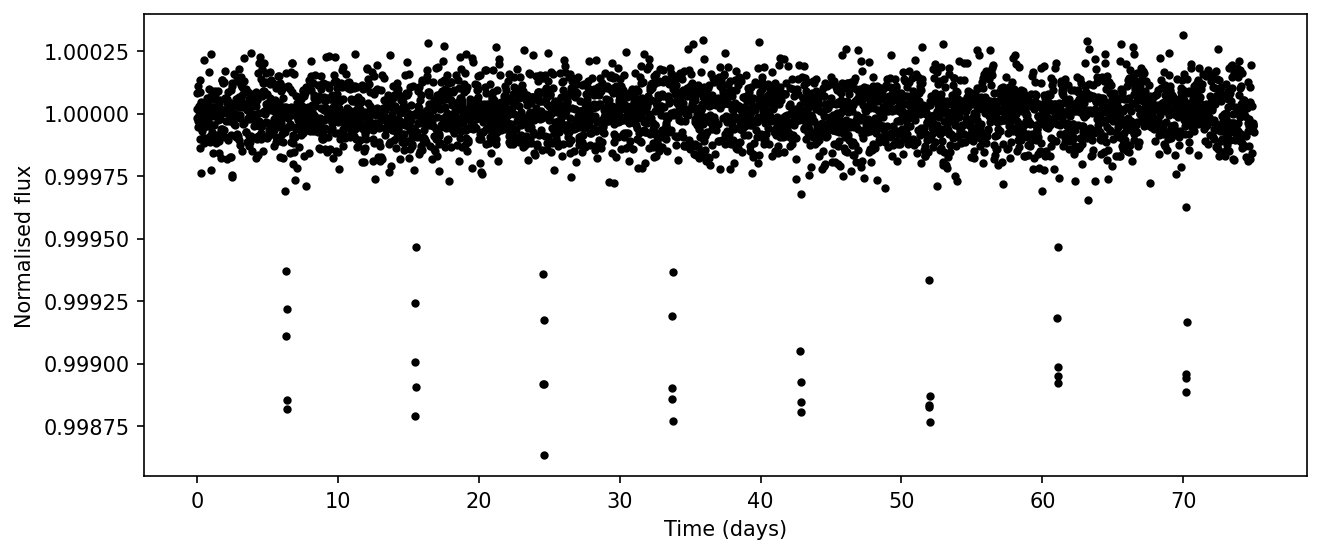}
\caption{Simulated light curve of host star of the planetary system of Table \ref{tab: data}.}

\label{fig: estrella_1_lc}
\end{figure}

\begin{figure}[H]
\includegraphics[width=13.5cm]{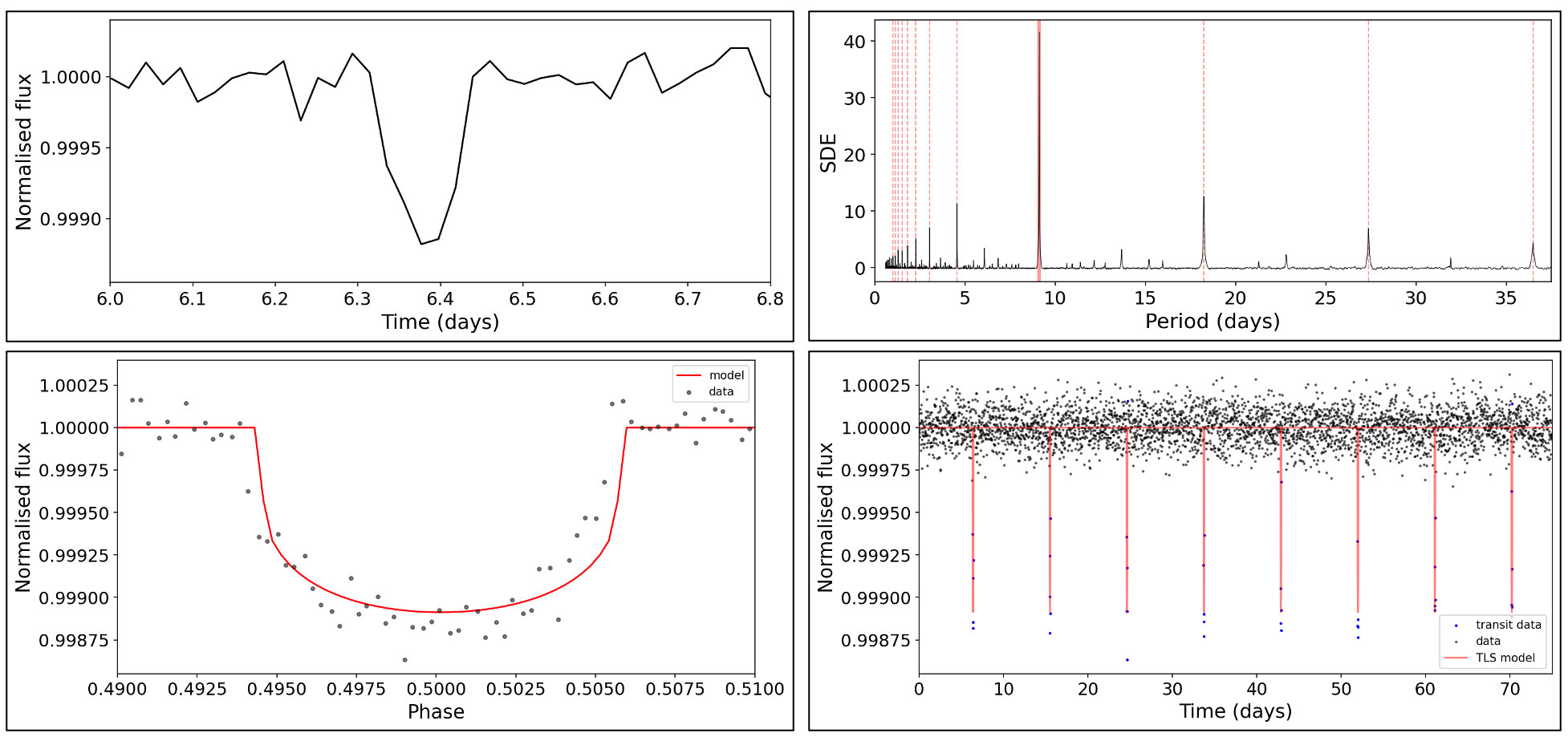}
\caption{\textls[-15]{Results of the analysis of the light curve shown in Figure \ref{fig: estrella_1_lc}. \textbf{Left column}: In the top panel, magnification of the epoch. In the bottom panel, phase-folded light curve (black dots) and the computed Mandel and Agol shape with TLS (red line). \textbf{Right column}: In the top panel, SDE graphic, where the highest peak (in red) corresponds to the orbital period. In the bottom panel, the light curve (black dots) with the detected transits by TLS (blue dots) and the computed non-phase-folded transit model (red line).}}
\label{fig: Estrella_1}
\end{figure}

We conducted a real TLS~\citep{TLS} analysis of one of the simulated light curves to check if the simulated transits could be detected with traditional analysis. These planetary system parameters are shown in Table \ref{tab: data}, and the simulated light curve is shown in Figure~\ref{fig: estrella_1_lc}. We~magnified the epoch of this light curve for testing the Mandel and Agol shape (top left panel in Figure \ref{fig: Estrella_1}) to verify its shape. As~shown, the transit is affected by noise, so it does not follow a perfect Mandel and Agol shape (see Figure \ref{fig: model}), which is common in real light curve data. After analyzing the light curve (Figure \ref{fig: estrella_1_lc}) with TLS, we obtained that it presented a transit-like signal, as shown in the top right panel in Figure \ref{fig: Estrella_1}, which corresponds to the signal detection efficiency (SDE) graphic, which plots the likelihood of each possible value of the orbital period being the real one. If one clear peak exists in this graphic (as in this case), the code has found a possible transit-like signal in the light curve. Additionally, we computed the phase-folded light curve (bottom left panel in Figure~\ref{fig: Estrella_1}). As~shown, the computed Mandel and Agol shape fits the data well (black dots). In~addition, TLS allowed us to plot which of the points of the light curve were selected as part of the transits. This~result is shown in the bottom right panel in Figure \ref{fig: Estrella_1}. Again, black dots correspond to the simulated light curve, blue dots are the light curve transit points, 
and the red line shows the computed model (without folding it upon the epoch). All these results show that our simulated light curves with transits could be detected with traditional models.

\begin{table}[H]
\caption{Main parameters used to simulate one of the light curves of the training dataset. We~used this simulated planetary system to check our simulations.}
\label{tab: data}
\newcolumntype{C}{>{\centering\arraybackslash}X}
\begin{tabularx}{\textwidth}{CCCCCCCC}
\toprule
\textbf{P(Days)} & \boldmath{$t_0$}\textbf{(Days)} & \boldmath{$R_p[R_\star]$} & \boldmath{$a[R_\star]$} & \textbf{i{[}deg{]}} & \textbf{mag}  & \boldmath{$R_\star[R_\odot]$} & \boldmath{$M_\star[M_\odot]$} \\ \midrule
9.12  & 6.38    & 0.03      & 27.09    & 90.00    & 14.86 & 1.00        & 3.21        \\ \bottomrule
\end{tabularx}

\end{table}
\unskip
\unskip

\begin{figure}[H]
\includegraphics[width=\linewidth]{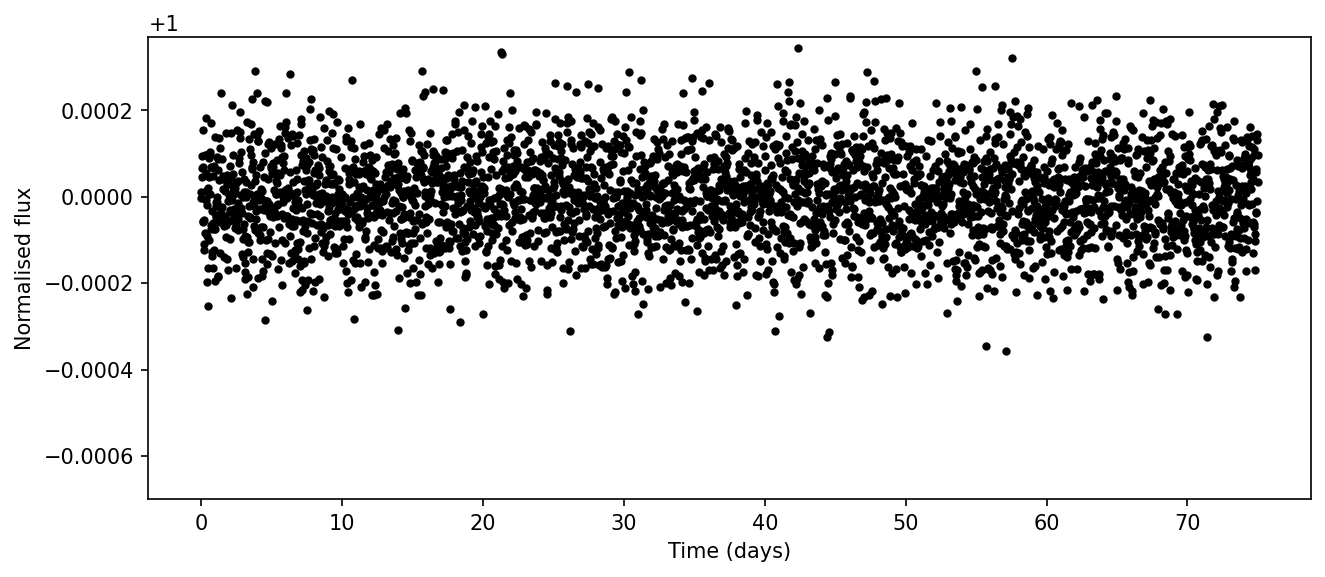}
\caption{Simulated light curve without transits.}
\label{fig: estrella_2_lc}
\end{figure}

\begin{figure}[H]
\includegraphics[width=\linewidth]{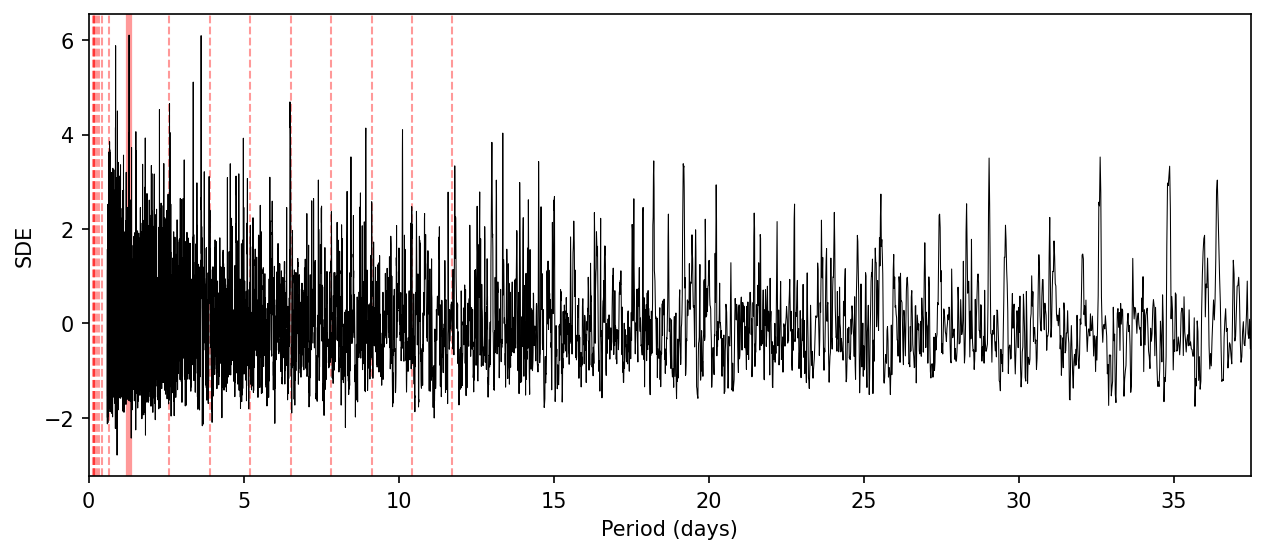}
\caption{SDE from the analysis with TLS. No clear peaks are shown, which means that it does not present transit-like signals.}
\label{fig: SDE_estrella_2}
\end{figure}

Not all the simulated light curves should have transits because we aimed to create a CNN that is able to detect transits, so we also needed light curves without transits. For~this aim, we used a Gaussian distribution to compute a pure noise light curve. This~was the same as in the light curves with transits, but we changed the median value of the Gaussian to 1, and we did not use batman. We~clipped the light curves as in the previous case, thus preventing the CNN from learning to distinguish between both types by analyzing the presence (or not) of outliers. A light curve without transits is shown in Figure \ref{fig: estrella_2_lc}. We~also analyzed it with TLS, finding that no outstanding peaks appeared in the SDE graphic (Figure \ref{fig: SDE_estrella_2}), confirming that the light curve did not present transit-like signals.

\unskip

We repeated this analysis for a set of 50,000 light curves from the test dataset but  using BLS, which is much faster. We~obtained a 99.5\% accuracy. This~helped to confirm that our simulated light curve transits could be detected with traditional analysis (which means that they are sufficiently realistic). Additionally, the time required to analyze each light curve with TLS was approximately 30 s and about 1--2 s with BLS, which was much faster with our neural network model (see Section \ref{sec: train_results}). 

We created $1.3\cdot 10^6$ light curves for our training dataset and $3\cdot 10^5$ for the testing dataset. In each, approximately 50\% of the light curves had transits, and the other 50\% did not. This~is because although many more light curves without detectable transits exist in K2 light curves, if we wanted the CNN to learn to detect transit-like signals, the amount of transit-like signals in the datasets needed to be in the order of half of the total amount of light curves, regardless of the fact that fewer light curves  actually present transit-like signals. After simulating all the light curves from the training dataset, all the flux data (because the times are not relevant in 1D CNN) were saved in one vector, and we created another one with the same length that has the label of each light curve, i.e., a value of 1 if the light curve had transits and 0 if it did not. The~same process was performed with the test dataset.

For our test, we needed to use a large enough light curve dataset (in the order of a few hundred thousand) to verify that dependency on the training dataset was not generated. As~the K2 mission only provides $\sim$1500 light curves with transit-like signals, we decided to use synthetic light curves for the test dataset.

\section{Training, Results, and Discussion}\label{sec: train_results}
As a CNN applies filters to the data from which it picks the highest value, and that data should be normalized before passing them to the CNN, we applied the following transformation to the light curves from the training and testing datasets. Let us refer to the flux vectors as fluxes. We~applied the following transformation for both datasets:

\begin{equation}
fluxes= 1-\frac{fluxes - min(fluxes)}{max(fluxes)-min(fluxes)}
\label{eqn: transform}
\end{equation}

As this algorithm was implemented in Python, we separately applied it to each of the elements present in the vector fluxes. The~result was that all the fluxes in the vector were inverted and reshaped between 0  and 1. This~was due to the convolutional kernels deleting the transits while selecting the maximum value where they were applied. This~did not affect  the shape of the transits but changed the way they could be understood. For~our 1D CNN model, transits are increases in the flux of light curves.

Before training our model, we had to compile it. We~decided to use Adam (adaptive moments)~\citep{2014arXiv1412.6980K} as a compiler, which is a stochastic gradient descent (SGD) method~\citep{10.1214/aoms/1177729586} that computes adaptive learning rates by using the values obtained from the first- and second-order gradients. Adam allows changes in the initial value of the learning rate (by default 0.001), so we used 0.00001. Furthermore, we used the accuracy as an evaluation metric and the binary cross-entropy as a loss function~\citep{doi:10.1198/016214506000001437} (entropy is a measure of the uncertainty related to a given distribution $q(y)$):

\begin{equation}
H_p(q)=\frac{1}{N}\sum_{i=1}^N y_i\cdot log(p(y_i))+(1+y_i)\cdot log(1-p(y_i))
\end{equation}
where $i=1,...,N$ is the number of datapoints, $y_i$ is the class associated with point $i$, and $p(y_i)$ is the predicted probability of point $i$ of being in class 1 (as we had two classes: transit~= 1 and not transit~= 0. $p(y_i)$ is the probability of having transits, and ($1-p(y_i)$) is the probability of not having them).

As we had a large training dataset, we decided to split it to perform the validation of the training. Keras allows the splitting of a validation set from the training set by selecting a number between 0 and 1 that corresponds to the percentage used to create the dataset (in our case, we set it to 0.3, which corresponds to 30\% of the training dataset). For~monitoring the training, we chose validation loss, and we defined two callbacks. The~first one, which is known as EarlyStopping, allows stopping the training if the validation loss starts to increase during a defined number of epochs (in our case, we choose two epochs); the second one, which is known as ModelCheckpoint, saves the best training results obtained. This~allowed us to obtain the best results where weights were included, although the models suffered from overfitting in the last two epochs (where the callbacks stopped the training). As~we defined these two callbacks, we could select a higher or lower number of epochs, so we decided to select 50 training epochs. In addition, we chose a batch size of 16, which corresponded to the number of samples that were propagated through the neural network.

\begin{figure}[H]
\begin{adjustwidth}{-\extralength}{0cm}
\includegraphics[width=18.5cm]{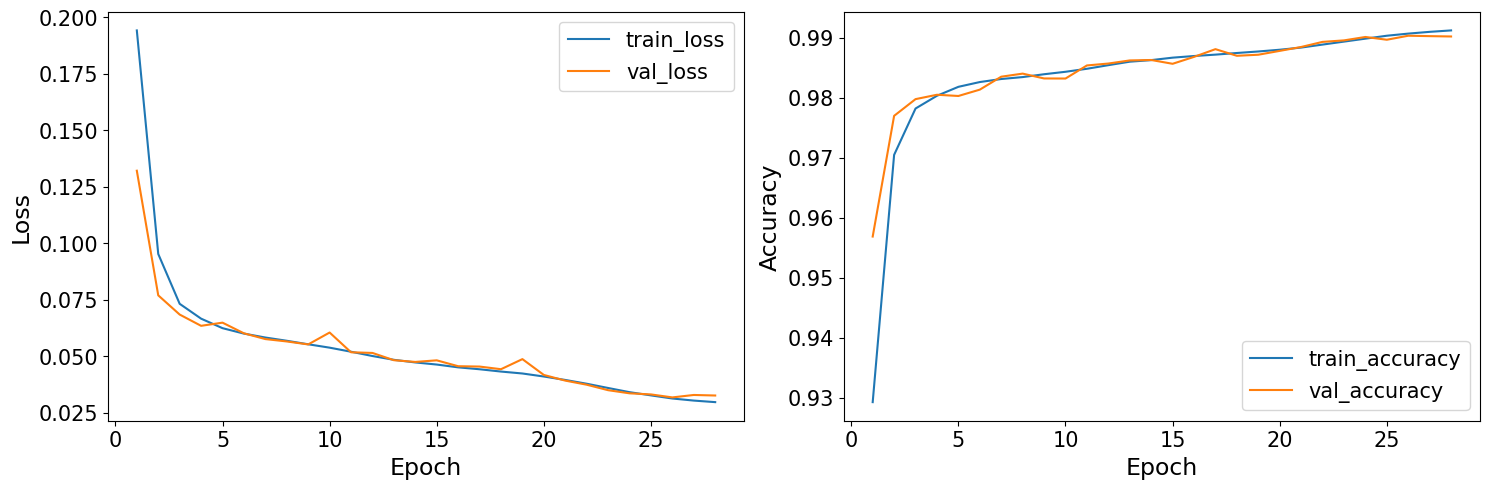}
\end{adjustwidth}
\caption{Training history of the CNN model. In the left panel, the loss function against the epochs for the training (blue) and validation (orange). In the right panel, the accuracy as a function of the epochs for the training (blue) and validation (orange). }
\label{fig: history}

\end{figure}

After training the model, we plotted the training history in which the validation loss, validation accuracy, training loss, and training accuracy were plotted against the epochs (Figure \ref{fig: history}). A detailed inspection of this 
figure shows that the training developed properly.

The model training was conducted on a server with an Intel Xeon E4-1650 V3, 3.50 GHz, with 12 CPUs (6 physical and 6 virtual). It~had a RAM of 62.8 Gb. Each epoch required approximately 1060 s (approximately 17.66 min), which means that the entire training of the model required approximately 8.83 h.

\begin{figure}[H]
\includegraphics[width=9.5 cm]{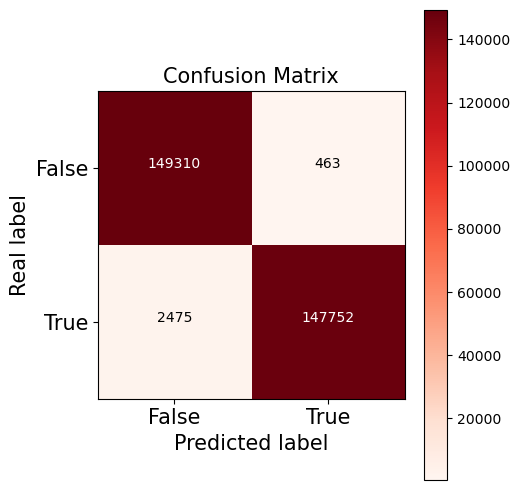}
\caption{Confusion matrix obtained from predicting on the test dataset.}
\label{fig: confusion_matrix}
\end{figure}

With the test dataset, we performed the model test. This~helped us to find out if the model generalized correctly or if it generated dependency on the training dataset. By~creating a confusion matrix, which compared the predicted labels for the model test and the real ones, we could graphically check the results (Figure \ref{fig: confusion_matrix}).

\begin{figure}[H]
  
  \includegraphics[width=0.8\linewidth]{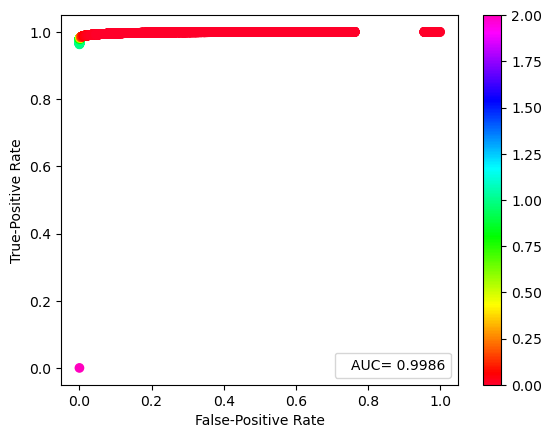}
  \caption{True positive rate as a function of false positive rate (ROC curve) scaled considering the threshold used to compute the MISSING 
  The probabilities should be lower than 1.0. The~pink dot (with a value of 2.0) does not represent predictions because it is an arbitrary value used as a starting point for the process. We~also plot the value of the computed AUC.}
  \label{fig: AUC}
\end{figure}

\textls[-15]{In this test of the model, we obtained an accuracy of 99.02\% and an estimated error (loss function) of 0.03. Additionally, as our dataset was balanced (meaning that we had $50 \%$ of each light curve type), and we did not want our classifier to be dependent on being trained with the same amount of light curves with and without transits, we also used the true skill statistic~\citep{2012ApJ...747L..41B}, which is independent of the balance of the training dataset. It~is defined as the difference between the true positive rate (TPR) and the false positive rate (FPR): }

\begin{equation}
TSS:= TPR-FPR= \frac{TP}{TP+FN}-\frac{FP}{TN+FP}
\label{eqn: TSS}
\end{equation}
\textls[-15]{where TP, TN, FN, and FP are, respectively, the number of true positives, true negatives, false negatives, and false positives obtained during the testing of the model, as shown in Figure \ref{fig: confusion_matrix}. This~parameter can vary between --1 and +1, where --1 means that the classifier always makes incorrect predictions and +1 means that it is always correct in its predictions. We~obtained a value of $0.97$, which means that the model correctly predicted most of the curve types. In addition, we used the area under the curve (AUC) of the receiver operating characteristic (ROC) curve, which compares the TPR and the FPR, when moving the threshold of the probability domain of each class, which is not given by the classifier. The~AUC represents the probability of a randomly positive value being ranked higher by the classifier than a randomly chosen negative one. The~area is computed by using a trapezoidal rule. The~results obtained are shown in Figure \ref{fig: AUC}. In an ideal classifier, where TPR~=~1.0 and FPR~=~0.0, the ROC curve should reach this point (0, 1) and its AUC should be 1.}

We obtained AUC = 0.99, and the ROC curve reached a point really close to (0, 1), which means that our classifier behaves correctly.

All these results showed that the model correctly generalizes and is able to classify our simulated light curves, which were previously checked with current analysis (i.e., BLS and TLS), even though it was not trained with them. This~is a promising result because, as this model learns from simulated light curves, it can probably be generalized to real K2 light curves, allowing the detection of unknown planetary system transits and increasing the statistics related to the amount of candidate planetary systems. Notably, for confirming a planetary candidate, their mass must be estimated, something that cannot be achieved with the transit technique. Instead, the most common method used is the radial velocities technique, which consists of measuring the Doppler shifts in the host star spectra due to the gravitational interaction between the star and planet. With the result obtained, we also confirm that the 1D CNN is a suitable choice for working with non-phased-folded Mandel and Agol light curves with transits, which provides an extra contribution to the literature~\citep{2018MNRAS.474..478P, Zucker_2018}.

\textls[-15]{To compare the performance of our method with that of traditional transit search methods, such as box least squares (BLS), we analyzed a set of 50,000 light curves with BLS (see Section~\ref{sec: train_test}). The~BLS search resulted in a 99.5\% detection accuracy, which is slightly higher than that of our 1D CNN model (99.02\%), but the 1D CNN search was completed in approximately 0.1\% of the time of the BLS search (74 s for the 1D CNN versus 75,000 s for BLS and on the same set of 50,000 light curves). The~small difference in the accuracy between the methods is compensated for by the huge reduction in computational time.}

\section{Conclusions}\label{sec: conclusions}
In this paper, we presented a 1D convolutional neural network model that was trained, validated, and tested with simulated Mandel and Agol~\citep{Mandel_2002} light curves with and without transits to create a convolutional classifier that could be generalized to real K2 data. We~also created a light curve simulator that could be easily adapted to create light curves from different surveys by changing the time domain of the data and the data collection cadence. It~also allows for the simulation of a wide range of planetary systems in which we can control the noise levels to complicate the detections. The~results obtained are highly satisfactory because we obtained high accuracy during the training, validation, and testing of the model (accuracy of 99.02\%). In this study, we applied a different approach compared with that in previous studies~\cite{2018MNRAS.474..478P, Zucker_2018}; we confirm, as have other researchers~\cite{2018MNRAS.474..478P, Zucker_2018, Ansdell_2018, 2019MNRAS.488.5232C, Shallue_2018}, that convolutional neural networks seem to be one of the best machine learning techniques for solving the principal problems related to the transits technique, which are the amount of time required to visually inspect all the light curves obtained from a survey and to analyze them and the number of light curves that have to be visually inspected by researchers to decide if a light curve presents transit-like signals. Future studies will be aimed at generalizing our model to real data and examining the possibility of estimating different planetary parameters that are related to a Mandel and Agol transit shape, such as the ones shown in Section \ref{sec: train_test}. We~will also simulate transit-like signals that are not produced by exoplanets, such as eclipsing binary light curves, which are common false-positive signals that can confuse a traditional transit analysis.

 \vspace{6pt}

\authorcontributions{\textls[-15]{Research: S.I.Á.; Coding: S.I.Á. and E.D.A.; Writing: S.I.Á.; Reviewing and Editing: E.D.A., M.L.S.R., J.R.R. and F.J.d.C.J. Formal Analysis: F.S.L. and M.L.S.R.; Manuscript Structure: S.I.Á., E.D.A. and J.R.R. All authors have read and agreed to the published version of the manuscript.}}

\funding{This reseach was funded by Proyecto Plan Regional by FUNDACION PARA LA INVESTIGACION CIENTIFICA Y TECNICA FICYT, grant number SV-PA-21-AYUD/2021/51301 and Plan Nacional by Ministerio de Ciencia, Innovación y Universidades, Spain, grant number MCIU-22-PID2021-127331NB-I00.}

\acknowledgments{\textls[-15]{We used the NASA Exoplanet Archive, which is operated by the California Institute of Technology, under contract with the National Aeronautics and Space Administration under the Exoplanet Exploration Program. This~research has made use of NASA’s Astrophysics Data System Bibliographic Services. We~are grateful for the prior work carried out by Lidia Sainz Ledo. The~authors would like to thank Carlos González Gutiérrez for his valuable comments and suggestions.}}

\conflictsofinterest{The authors declare no conflicts of interest.} 




\begin{adjustwidth}{-\extralength}{0cm}
\reftitle{References}

\PublishersNote{}
\end{adjustwidth}
\end{document}